 \UseRawInputEncoding 
\documentclass[12pt]{article}
\usepackage{amssymb}
\usepackage{amsmath}
\usepackage{enumerate}
\usepackage{graphicx}
\usepackage{epstopdf}
\usepackage{cite}
\usepackage{makecell}
\newtheorem{theorem}{Theorem}

\newtheorem{lemma}{Lemma}

\newtheorem{Exa}{Example}
\newtheorem{corollary}{Corollary}

\input amssym.def
\newcommand\btd{\raise 2pt \hbox{$\hat\bigtriangledown$}\hskip 1.5pt}
\newcommand\bt{\raise 2pt \hbox{$\bigtriangledown$}\hskip 1.5pt}
\usepackage{indentfirst}
\newcommand{\ket}[1]{\left|#1\right>}

\usepackage{graphicx}
\usepackage[justification=centering]{caption}

\usepackage{color}

\begin{document}
\title{Multipartite concurrence of $W$-class states based on sub-partite quantum systems}
\author{\normalsize Wei Chen$^{1}$, Yanmin Yang$^{2}$ $\thanks{e-mail: ym.yang@kust.edu.cn}$, Shao-Ming Fei$^{3,4}$,  Zhu-Jun Zheng$^{5}$, Yan-Ling  Wang$^{1}$\\
{\footnotesize $^{1}$ School of Computer Science and Technology, }\\
{\footnotesize Dongguan University of Technology, Dongguan, 523808, P.R. China}\\%[.3cm]
{\footnotesize $^{2}$ Faculty of Science, }\\
{\footnotesize Kunming University of Science and Technology, Kunming,  650500, P.R. China}\\
{\footnotesize $^{3}$ School of Mathematical Sciences, }\\
{\footnotesize Capital Normal University, Beijing 100048,  P.R. China}\\
{\footnotesize $^{4}$ Max-Planck-Institute for Mathematics in the Sciences, Leipzig 04103, Germany}\\
{\footnotesize $^{5}$ School of Mathematics, }\\
{\footnotesize South China University of Technology, Guangzhou 510641, P.R. China}\\
}
\date{}
\maketitle

\begin{abstract}
We study the concurrence for arbitrary $N$-partite $W$-class states based on the $(N-1)$-partite partitions of subsystems by taking account to the structures of $W$-class states. By using the method of permutation and combination we give analytical formula of concurrence and some elegant relations between the multipartite concurrence and the $(N-1)$-partite concurrence for arbitrary multipartite $W$-class states. Applying these relations we present better lower bounds of concurrence for multipartite mixed states. An example is given to demonstrate that our lower bounds can detect more entanglements.
\medskip

\textbf{Keywords}{~~Multipartite concurrence $\cdot$ \and W-class states $\cdot$ \and $(N-1)$-partite partitions $\cdot$ \and Lower bound of concurrence }
 %\PACS{03.67.Mn \and 03.65.Ud }
\end{abstract}

\section{Introduction}\label{sec1}
\label{intro}
Quantum entanglement is a striking feature of quantum physics \cite{Nie,Ein,Ami,Wer} and an essential resource in quantum information processing varying from quantum teleportation \cite{Bennett} and quantum cryptography \cite{Ekert} to dense coding
\cite{Bennett1}. Due to its variety of usages, quantum entanglement has attracted much attention in recent years \cite{Hor,Flo,Chen,Bre,Bre1,Vic,Zhang}.

To quantify the entanglement of a state, the concept of entanglement measure has been naturally introduced, such as the entanglement of formation \cite{Ben2} for bipartite quantum systems and concurrence \cite{Uhl} for any multipartite quantum systems.
For the two-qubit case, the entanglement of formation is proven to be a monotonically increasing function of the concurrence and an
elegant formula for concurrence was derived analytically by Wootters \cite{Woo}. However, except for bipartite qubit systems and some special symmetric states \cite{Ter}, there have been no explicit analytic formulas of concurrence for arbitrary high-dimensional mixed states,
due to the extremizations involved in the computation.

Instead of analytic formulas, some progress has been made toward the analytical lower bounds of concurrence. In \cite{Fan,Zhu}, the authors presented a lower bound of concurrence by
decomposing the joint Hilbert space into many $2\otimes 2$ and $s\otimes t$-dimensional subspaces, which improve all the known lower bounds of concurrence. Similar nice algorithms and progress have been made towards the lower bounds of concurrence for tripartite quantum systems \cite{Zhu0,Ch} and other multipartite quantum systems \cite{Wang,Zhu1} based on bipartite partitions of the whole quantum system. The authors in \cite{Ch1} improve the lower bound of concurrence by using tripartite and $M$-partite concurrences of an $N$-partite ($2\leq M < N$) systems.

As a particular kind of quantum states, the well-known $W$-class states \cite{Kim1} have been widely studied. In this paper we first study the multipartite concurrence for $W$-class states and derive an analytical formula for pure $W$-class states. Then we investigate the $N$-partite concurrence of $W$-class states based on the $(N-1)$-partite quantum systems and present an elegant relation between among them. Based on the results for the $W$-class states, we derive better lower bounds of concurrence for a class of multipartite mixed states. An example is given to illustrate that our lower bound may detect more entanglements.
\medskip

\section{Multipartite concurrence of $W$-class states}\label{sec2}
We first recall the definition of the multipartite concurrence.
Let $H_i$, $i=1,\cdots,N$, be $d_i$ dimensional Hilbert spaces.
The concurrence of an $N$-partite pure state $|\psi\rangle\in H_1\otimes H_2\otimes\cdots \otimes H_N$ is defined by \cite{Aol},
\begin{equation}\label{1}
\mathcal{C}_N(|\psi\rangle) = 2^{1-\frac{N}{2}}\sqrt{(2^N-2)-\sum_{\alpha} Tr(\rho_{\alpha}^{2})},
\end{equation}
where the index $\alpha$ labels all $2^N-2$ non-trivial subsystems of the $N$-partite quantum systems and $\rho_{\alpha}$ are the corresponding reduced density matrices.
For a mixed multipartite quantum state $\rho = \sum_{i}p_{i}|\psi_{i}\rangle\langle\psi_{i}| \in H_1\otimes H_2\otimes\cdot\cdot\cdot \otimes H_N$,
$p_{i}\geq 0$, $\sum_{i}p_{i} = 1$, the concurrence is given by the convex roof:
\begin{equation}\label{2}
\mathcal{C}_N(\rho) = \min_{\{p_{i},|\psi_{i}>\}}\sum_{i}p_{i}\mathcal{C}_N(|\psi_{i}\rangle),
\end{equation}
where the minimum is taken over all possible pure state decompositions of $\rho$.

In \cite{Wang} the authors obtained the lower bounds of multipartite concurrence in terms of the concurrences of bipartite partitioned states of the whole quantum system. For an $N$-partite quantum pure state $|\psi\rangle\in H_1\otimes H_2\otimes\cdots \otimes H_N$,
the concurrence of bipartite partition between the subsystems $1 2 \cdots M$ and $M+1\cdots N$ is defined by
\begin{eqnarray}\label{3}
\mathcal{C}_2(|\psi\rangle\langle\psi|) = \sqrt{2(1-Tr(\rho^{2}_{1 2\cdots M}))},
\end{eqnarray}
where $\rho_{1 2\cdots M} = Tr_{M+1\cdots N}\{|\psi\rangle\langle\psi|\}$
is the reduced density matrix of $\rho = |\psi\rangle\langle\psi|$ by tracing over the subsystems $M+1\cdots N.$
For a mixed multipartite quantum state $\rho = \sum_{i}p_i |\psi_{i}\rangle\langle\psi_{i}| \in  H_1\otimes H_2\otimes\cdots \otimes H_N,$
the corresponding concurrence $\mathcal{C}_2(\rho)$ is given by the convex roof:
\begin{eqnarray}\label{4}
\mathcal{C}_2(\rho) = \min_{\{p_{i},|\psi_{i}\rangle\}}\sum_{i}p_{i}\mathcal{C}_2(|\psi_{i}\rangle\langle\psi_{i}|).
\end{eqnarray}
A relation between the concurrence (\ref{2}) and the bipartite concurrence (\ref{4}) has been presented in \cite{Wang}.
For a multipartite quantum state $\rho \in  H_1\otimes H_2\otimes\cdots \otimes H_N$ with $N\geq 3,$ the following inequality holds,
\begin{eqnarray}\label{5}
\mathcal{C}_N(\rho) \geq \max 2^{\frac{3-N}{2}}\mathcal{C}_2(\rho),
\end{eqnarray}
where the maximum is taken over all kinds of bipartite concurrences.

In terms of the lower bounds of bipartite concurrence, in \cite{Zhu1}
further relations between the concurrence (\ref{2}) and the bipartite concurrence (\ref{4}) have been derived:
\begin{eqnarray}\label{6}
\mathcal{C}_N(\rho) \geq \max_{M = 1, 2, \cdots, N-1} \{2^{\frac{1-N}{2}}\sqrt{2^{N-M}+2^M-2}\mathcal{C}_2(\rho_M)\}
\end{eqnarray}
for $N\geq 3,$ where the maximum is taken over all kinds of bipartite concurrences for given $M$. In particularly, if $N=3,$ one has $\mathcal{C}_3(\rho) \geq \max \{\mathcal{C}_2(\rho_1), \mathcal{C}_2(\rho_2)\}$. If $N=4,$ one gets $\mathcal{C}_4(\rho) \geq \max \{\mathcal{C}_2(\rho_1), \frac{\sqrt{3}}{2}\mathcal{C}_2(\rho_2), \mathcal{C}_2(\rho_3)\}.$

In order to improve the lower bounds of concurrence, instead of the bipartite concurrence $\mathcal{C}_{2}(\rho)$, the authors in \cite{Ch} consider tripartite concurrence $\mathcal{C}_{3}(\rho)$. In \cite{Ch1} the authors improve the lower bound of concurrence by using tripartite and $M$-partite concurrences of an $N$-partite ($M < N$) system.
For an $N$-partite quantum pure state $|\psi\rangle \in H_1\otimes H_2\otimes\cdots \otimes H_N$ $(N\geq 3)$, denote $\{i^1\}$, $\{i^2\}$, $\cdots$, $\{i^{M_1}\}$, $\{k_1^1, k_2^1\}$, $\{k_1^2, k_2^2\}$, $\cdots$, $\{k_1^{M_2}$, $k_2^{M_2}\}$, $\cdots$, $\{q_1^{1}, \cdots, q_j^{1}\}$, $\{q_1^{2},
\cdots, q_j^{2}\}$, $\cdots$, $\{q_1^{M_j}, \cdots, q_j^{M_j}\}$ as the $M$ decompositions among the subsystems, where $\{i^1, i^2, \cdots,   i^{M_1},  ~k_1^1, ~k_2^1,  ~k_1^2, ~k_2^2, \cdots, k_1^{M_2}, k_2^{M_2},\cdots, q_1^{1},\cdots, q_j^{1}, \cdots, q_1^{M_j},\cdots, q_j^{M_j}\} = \{1,2,\cdots,N\}$ and $\sum_{k=1}^{j}M_{k}=M,$ $\sum_{k=1}^{j}kM_{k}=N$. The concurrence of the $M$-partite decompositions among the above subsysytems is given by
\begin{eqnarray}\label{M-partite}
\mathcal{C}_{M}(|\psi\rangle\langle\psi|) = 2^{1-\frac{M}{2}}\sqrt{(2^M-2)-\sum_{\alpha}Tr(\rho_{\alpha}^{2})},
\end{eqnarray}
where $\emptyset \neq \alpha \subsetneq \{\{i^1\}, \{i^2\}, \cdots,\{i^{M_1}\}, \{k_1^1, k_2^1\}, \{k_1^2, k_2^2\},$ $ \cdots,\{k_1^{M_2}, k_2^{M_2}\},\cdots,\{q_1^{1},\cdots, q_j^{1}\},\cdots,$
\noindent$ \{q_1^{M_j},\cdots, q_j^{M_j}\}\}$ and $\rho_{\alpha}$ are the corresponding reduced density matrices. The rearrangement of the subsystems are implied naturally. Taking $N=4$ and $M=3,$ one has six different partitions of the four-partite system: $1|2|34, 1|3|24, 1|4|23, 12|3|4, 13|2|4$ and $14|2|3.$
In terms of the lower bounds of tripartite concurrence, in \cite{Ch1} the authors derived a relation between the concurrence (\ref{2}) and the bipartite concurrence (\ref{M-partite}), $C_4^{2}(\rho) \geq  \widetilde{C_3}^{2} (\rho),$ where $\widetilde{C_3}^{2} (\rho) = \frac{1}{6}(C_3^{2}(\rho_{1|2|34})+C_3^{2}(\rho_{1|3|24})
+C_3^{2}(\rho_{1|4|23})+C_3^{2}(\rho_{12|3|4})+C_3^{2}(\rho_{13|2|4})+C_3^{2}(\rho_{14|2|3})).$

In order to improve the above lower bounds, we first consider the $N$-qubit $W$-class states \cite{Kim1},
\begin{eqnarray}\label{w-state}
{|W\rangle}_{A_1 A_2 \cdots A_N} = a_1 |10\cdots 0\rangle_{A_1 A_2 \cdots A_N} +a_2 |01\cdots 0\rangle_{A_1 A_2 \cdots A_N} + \cdots +a_N |00\cdots 1\rangle_{A_1 A_2 \cdots A_N},
\end{eqnarray}
where  $\sum_{i=1}^{N}|a_i|^2=1$. Let $\rho \triangleq\rho_{A_1 A_2 \cdots A_N} = |W\rangle_{A_1 A_2 \cdots A_N}\langle W|$ and
\begin{eqnarray}
\nonumber
  \rho_{{i_1}{i_2}\cdots {i_k}} \triangleq \rho_{A_{i_1}A_{i_2}\cdots A_{i_k}}
  = tr_{A_1 \cdots \widehat{A_{i_1}}\cdots \widehat{A_{i_2}} \cdots \widehat{A_{i_k}}\cdots A_N}(|W\rangle_{A_1 A_2 \cdots A_N}\langle W|)
\end{eqnarray}
for any $1\leq i_1<i_2<\cdots<i_k \leq N$. We have
\begin{eqnarray}\label{rho}
\nonumber
  \rho_{{i_1}{i_2}\cdots {i_k}}
  = (a_{i_1}|10\cdots0\rangle+\cdots+a_{i_k}|00\cdots1\rangle)_{A_{i_1}A_{i_2}\cdots A_{i_k}}(a_{i_1}^*\langle10\cdots0|+a_{i_k}^*\langle00\cdots1|)\\
  +\sum_{j\neq i_1,\cdots,i_k}|a_j|^2|00\cdots0\rangle_{A_1 \cdots \widehat{A_{i_1}}\cdots \widehat{A_{i_2}} \cdots \widehat{A_{i_k}}\cdots A_N}\langle00\cdots0|,
\end{eqnarray}
and
\begin{eqnarray}\label{1-Tr}
  1-Tr(\rho^2_{{i_1}{i_2}\cdots {i_k}})
  =2(|a_{i_1}|^2+\cdots+|a_{i_k}|^2)\sum_{j\neq i_1,\cdots,i_k}|a_j|^2,
\end{eqnarray}
where $A_1 A_2 \cdots \widehat{A_i} \cdots A_N = A_1 A_2 \cdots A_{i-1} A_{i+1} \cdots A_N.$
For simplicity, we denote by $\rho_{{i_1}{i_2}\cdots {i_k}}$ the reduced density operator $\rho_{A_{i_1}A_{i_2}\cdots A_{i_k}}$. We have the following lemmas for the $N$-qubit $W$-class states.

\begin{lemma}\label{lem1}
When $N$ ($N>1$) is even, we have
\begin{eqnarray}\label{even}
2C_{N}^0+2C_{N}^1+\cdots +2C_{N}^{\frac{N}{2}-1}+C_{N}^{\frac{N}{2}}=2^{N}.
\end{eqnarray}
\end{lemma}

\noindent{\bf Proof.} As $C_{N}^i= C_{N}^{N-i}$ for all integer $0\leq i< N$, we have
$$2C_{N}^0+2C_{N}^1+\cdots+2C_{N}^{\frac{N}{2}-1}+C_{N}^{\frac{N}{2}}~~~~~~~~~~~~~~~~~~~~~~~~~~~~~~~~~~~~~~~~~~~~~~~~$$
$$=C_{N}^0+C_{N}^1+\cdots+C_{N}^{\frac{N}{2}-1}+C_{N}^{\frac{N}{2}}+C_{N}^{\frac{N}{2}+1}+\cdots+C_{N}^{N}~~~~~~~~~~~~~~~~~~~~~~~$$
$$=(1+1)^{N}=2^{N}.~~~~~~~~~~~~~~~~~~~~~~~~~~~~~~~~~~~~~~~~~~~~~~~~~~~~~~~~~~~~~~~~~~~~~~~$$
Hence, $2C_{N}^0+2C_{N}^1+\cdots+2C_{N}^{\frac{N}{2}-1}+C_{N-2}^{\frac{N}{2}}=2^{N}.$ $\Box$

\begin{lemma}\label{lem2}
When $N$ ($N\geq 1$) is odd, we have
\begin{eqnarray}\label{odd}
2C_{N}^0+2C_{N}^1+\cdots +2C_{N}^{\frac{N-1}{2}}=2^{N}.
\end{eqnarray}
\end{lemma}

\noindent{\bf Proof.} As $C_{N}^i= C_{N}^{N-i}$ for all integer $0\leq i< N$, we have
$$2C_{N}^0+2C_{N}^1+\cdots +2C_{N}^{\frac{N-1}{2}}~~~~~~~~~~~~~~~~~~~~~~~~~~~~~~~~~~~~~~~~~~~~~~~~~~~~~~~~~~~~$$
$$=C_{N}^0+C_{N}^1+\cdots+C_{N}^{\frac{N-1}{2}}+C_{N}^{\frac{N+1}{2}}+C_{N-1}^{N}+C_{N}^{N}~~~~~~~~~~~~~~~~~~~~~~~~~~~~~~~~$$
$$=(1+1)^{N}=2^{N}.~~~~~~~~~~~~~~~~~~~~~~~~~~~~~~~~~~~~~~~~~~~~~~~~~~~~~~~~~~~~~~~~~~~~~~~~~~$$
Hence, $2C_{N}^0+2C_{N}^1+\cdots+2C_{N}^{\frac{N-1}{2}}=2^{N}.$ $\Box$

\begin{theorem}\label{Sec2-3}
The squared multipartite concurrence of the $N$-qubit $W$-class state ${|W\rangle}_{A_1 A_2 \cdots A_N}$ is given by
\begin{eqnarray}\label{theorem3}
\mathcal{C}_N^{2}({|W\rangle}_{A_1 A_2 \cdots A_N}) = 4 \sum\limits_{1\leq i<j\leq N}{|a_i|^2|a_j|^2}
\end{eqnarray}
for $N\geq 2$.
\end{theorem}

\noindent{\bf Proof.}
For $N =2$, we have ${|W\rangle}_{A_1 A_2} = a_1 |10\rangle_{A_1 A_2} +a_2 |01\rangle_{A_1 A_2}$ and $\mathcal{C}_2(|W\rangle_{A_1 A_2}) = \sqrt{(2^2-2)-Tr(\rho_{1}^{2})- Tr(\rho_{2}^{2})}$.  Then $\mathcal{C}_2^2(|W\rangle_{A_1 A_2}) = 2-Tr(\rho_{1}^{2})- Tr(\rho_{2}^{2}) = 4|a_1|^2|a_2|^2.$

i) $N>3$ and $N$ is even. Denote $d=\frac{N}{2}$. For pure states one has $ 1-Tr(\rho^2_{i_1~i_2\cdots i_k})= 1-Tr(\rho^2_{1 \cdots \widehat{i_1}\cdots \widehat{i_2} \cdots \widehat{i_k}\cdots N})$. By (\ref{1})
we have
$$\mathcal{C}_N^{2}({|W\rangle}_{A_1 A_2 \cdots A_N})=\frac{1}{2^{N-3}}[\sum\limits_{i=1}^N(1-tr\rho_{i}^2)+\sum\limits_{1 \leq i_1<i_2\leq N}(1-tr\rho_{i_1i_2}^2)+\cdots$$
\begin{eqnarray}\label{14}
+\sum\limits_{1\leq i_1<i_2<\cdots<i_{d-1}\leq N}(1-tr\rho_{i_1i_2\cdots i_{d-1}}^2)+\sum\limits_{1< i_1<i_2<\cdots<i_{d-1}\leq N}(1-tr\rho_{1i_1i_2\cdots i_{d-1}}^2)].
\end{eqnarray}
Concerning the numbers of the terms in the summations of (\ref{14}), we only need to consider the following non-trivial index $\alpha$, see \textbf{Table \ref{table1}},
\begin{table}[htb]
\centering
\begin{tabular}{|c|c|c|}
\hline
Type of $\alpha$ & Details of $\alpha$ & Number of $\alpha$  \\ \hline
with one subsystem &${1},{2}, \cdots, {N}$ & $C^{1}_{N}$  \\ \hline
with two subsystems &${12},{13}, \cdots, {(N-1)N}$ & $C^{2}_{N}$ \\ \hline
 $\cdots$ & $\cdots$ & $\cdots$ \\ \hline
with $d$ subsystems & ${12\cdots d},{13\cdots (d+1)}, \cdots, {1(d+2)\cdots N}$ & $C^{d-1}_{N-1}= \frac{1}{2}C^{d}_{N}$ \\ \hline
\end{tabular}
\textbf{\caption{Non-trivial index $\alpha$ in (\ref{14}).}\label{table1}}
\end{table}

By equality (\ref{1-Tr}), we have that each item of $1-Tr(\rho^2_{i_1i_2\cdots i_{d-1}})$  has the form $2|a_i|^2|a_j|^2.$ To compute $\mathcal{C}_N^{2}({|W\rangle}_{A_1 A_2 \cdots A_N})$ we just need to determine the total coefficients of the term $|a_i|^2|a_j|^2$ for every $1-Tr(\rho^2_{i_1i_2\cdots i_{d-1}})$ in (\ref{14}). Taking the coefficient of the term $|a_1|^2|a_2|^2$ as an example, see $\textbf{Table \ref{table2}}$, the total coefficient of the term $|a_1|^2|a_2|^2$ is $\frac{1}{2^{N-3}}(2\cdot 2C^{0}_{N-2}+2\cdot 2C^{1}_{N-2}+\cdots + 2\cdot C^{d-1}_{N-2})$, which is equal to 4 by \textbf{Lemma \ref{lem1}}. Similarly, we can prove that any item  $|a_i|^2|a_j|^2$ has the coefficient 4. Hence we have $\mathcal{C}_N^{2}({|W\rangle}_{A_1 A_2 \cdots A_N}) = 4 \sum\limits_{1\leq i<j\leq N}{|a_i|^2|a_j|^2}.$
\begin{table}[htb]\footnotesize
\centering
\begin{tabular}{|c|c|c|}
\hline
Type of $1-Tr(\rho^2_{i_1i_2\cdots i_{k}})$  & \thead{$1-Tr(\rho^2_{i_1i_2\cdots i_{k}})$\\ which has the  item $|a_1|^2|a_2|^2$} & The coefficient of $|a_1|^2|a_2|^2$  \\ \hline
$k=1$ & $1-tr(\rho^2_{1}),$ $1-tr(\rho^2_{2})$   & $2\cdot 2C^{0}_{N-2}$  \\ \hline
$k=2$ &\thead{ $1-tr(\rho^2_{13}), 1-tr(\rho^2_{14}), \cdots, $ $1-tr(\rho^2_{1N});$\\ $1-tr(\rho^2_{23}), 1-tr(\rho^2_{24}),\cdots, $ $1-tr(\rho^2_{2N})$ }& $2\cdot 2C^{1}_{N-2}$ \\ \hline
 $\cdots$ & $\cdots$ & $\cdots$ \\ \hline
$k=d-1$ & \thead{$1-tr(\rho^2_{134\cdots d}), \cdots, $ $1-tr(\rho^2_{1(d+3)\cdots N});$ \\ $1-tr(\rho^2_{234\cdots d}), \cdots, $ $1-tr(\rho^2_{2(d+3)\cdots N})$ }& $2\cdot 2C^{d-2}_{N-2}$ \\ \hline
$k=d$ & $1-tr(\rho^2_{134\cdots (d+1)}), \cdots,  $ $1-tr(\rho^2_{1(d+2)\cdots N})$ & $2\cdot 1 C^{d-1}_{N-2}$ \\ \hline
\end{tabular}
\centering
\textbf{\caption{The coefficients of the term $|a_1|^2|a_2|^2$ for even N).}\label{table2}}
\end{table}

ii) $N\geq3$ and $N$ is odd. Denote $d=\frac{N-1}{2}$. Similarly we have
$$\mathcal{C}_N^{2}({|W\rangle}_{A_1 A_2 \cdots A_N})=\frac{1}{2^{N-3}}[\sum\limits_{i=1}^N(1-tr\rho_{i}^2)+\sum\limits_{1 \leq i_1<i_2\leq N}(1-tr\rho_{i_1i_2}^2)+\cdots$$
\begin{eqnarray}\label{16}
+\sum\limits_{1\leq i_1<i_2<\cdots<i_{d-1}\leq N}(1-tr\rho_{i_1i_2\cdots i_{d-1}}^2)+\sum\limits_{1\leq i_1<i_2<\cdots<i_{d}\leq N}(1-tr\rho_{1i_1i_2\cdots i_{d-1}}^2)].
\end{eqnarray}
The non-trivial index $\alpha$ we need to consider is shown in \textbf{Table \ref{table3}}.
\begin{table}[htb]\footnotesize
\centering
\begin{tabular}{|c|c|c|}
\hline
Type of $\alpha$ & Details of $\alpha$ & Number of $\alpha$  \\ \hline
with one subsystem &${1},{2}, \cdots, {N}$ & $C^{1}_{N}$  \\ \hline
with two subsystems&${12},{13}, \cdots, {(N-1)N}$ & $C^{2}_{N}$ \\ \hline
 $\cdots$ & $\cdots$ & $\cdots$ \\ \hline
with $d$ subsystems  & ${12\cdots d},{13\cdots (d+1)}, \cdots, {1(d+3)\cdots N}, \cdots, {(d+2)(d+3)\cdots N}$ & $C^{d}_{N}$ \\ \hline
\end{tabular}
\textbf{\caption{Non-trivial index $\alpha$ in (\ref{16}).}\label{table3}}
\end{table}

In order to compute $\mathcal{C}_N^{2}({|W\rangle}_{A_1 A_2 \cdots A_N})$ we need to determine the total coefficient of the terms $|a_i|^2|a_j|^2$ for every $1-Tr(\rho^2_{i_1i_2\cdots i_{d-1}})$ in (\ref{16}). Still taking the coefficient of $|a_1|^2|a_2|^2$ as an example, from  $\textbf{Table \ref{table4}}$ we have that the total coefficient of the term $|a_1|^2|a_2|^2$ is $\frac{1}{2^{N-3}}(2\cdot 2C^{0}_{N-2}+2\cdot 2C^{1}_{N-2}+\cdots + 2\cdot 2C^{d-1}_{N-2})$ which is equal to $4$ by \textbf{Lemma \ref{lem2}}.
\begin{table}[htb]\small
\centering
\begin{tabular}{|c|c|c|}
\hline
Type of $1-Tr(\rho^2_{i_1i_2\cdots i_{k}})$  & \thead{$1-Tr(\rho^2_{i_1i_2\cdots i_{k}})$\\ which has the item $|a_1|^2|a_2|^2$} & The coefficient of $|a_1|^2|a_2|^2$  \\ \hline
$k=1$ & $1-tr(\rho^2_{1}),$ $1-tr(\rho^2_{2})$   & $2\cdot 2C^{0}_{N-2}$  \\ \hline
$k=2$ &\thead{ $1-tr(\rho^2_{13}), 1-tr(\rho^2_{14}), \cdots, $ $1-tr(\rho^2_{1N});$\\ $1-tr(\rho^2_{23}), 1-tr(\rho^2_{24}),\cdots, $ $1-tr(\rho^2_{2N})$ }& $2\cdot 2C^{1}_{N-2}$ \\ \hline
 $\cdots$ & $\cdots$ & $\cdots$ \\ \hline
$k=d$ & \thead{$1-tr(\rho^2_{134\cdots d}), \cdots, $ $1-tr(\rho^2_{1(d+3)\cdots N});$ \\ $1-tr(\rho^2_{234\cdots (d+1)}), \cdots, $ $1-tr(\rho^2_{2(d+3)\cdots N})$ }& $2\cdot 2C^{d-1}_{N-2}$ \\ \hline
\end{tabular}
\centering
\textbf{\caption{The coefficient of the term $|a_1|^2|a_2|^2$ (N odd).}\label{table4}}
\end{table}
Similarly, we can prove that any item  $|a_i|^2|a_j|^2$ has the coefficient $4$. Therefore, $\mathcal{C}_N^{2}({|W\rangle}_{A_1 A_2 \cdots A_N}) = 4 \sum\limits_{1\leq i<j\leq N}{|a_i|^2|a_j|^2}.$ $\Box$

\section{N-partite concurrence of $W$-Class States based on $(N-1)$-partite quantum systems}\label{sec4}

If $M=N-1$, under the rearrangement of the sub-systems there are $C_N^2$ different partitions of an $N$-partite system: $ij|1|\cdots|\widehat{i}|\cdots|\widehat{j}|\cdots|N$, $1\leq i<j\leq N$. Denote by $\widetilde{\mathcal{C}}_{N-1}^2(\ket{W_{A_1 A_2 \cdots A_N}})=\sum\limits_{\mathcal{P}}\mathcal{C}_{N-1}^2(\rho_{\mathcal{P}})$, where the index ${\mathcal{P}}$ labels all $C_N^2$ different $(N-1)$-partite partitions of the $N-$partite systems, and $\mathcal{C}_{N-1}^2(\rho_{\mathcal{P}})$ is the $(N-1)$-partite concurrence with respect to the partition $\mathcal{P}$.

\begin{theorem}\label{N-1 enev1}
For the $N$-qubit $W$-class states $\rho_{12 \cdots N} = |W\rangle_{A_1 A_2 \cdots A_N}\langle W|$, under the partition $12|3|\cdots|N$ we have
\begin{equation}\label{}
  \mathcal{C}_{N-1}^2(\rho_{12|3|\cdots|N})=4\sum\limits_{\substack{ 1\leq i<j\leq N\\(i,j)\neq (1,2)}}|a_i|^2|a_j|^2
\end{equation}
for $N\geq 3$.
\end{theorem}

\noindent{\bf Proof.}
i)  $N>3$ and $N$ is even. From equality (\ref{M-partite}) we have
\begin{equation}\label{sec4-1}
 \mathcal{C}^2_{N-1}(\rho_{12|3|\cdots|N}) = 2^{2-(N-1)}[(2^{N-1}-2)-\sum\limits_{\alpha} Tr(\rho_{\alpha}^{2})],
\end{equation}
where the index $\alpha$ labels all $2^{N-1}-2$ non-trivial subsystems of the $(N-1)$-partite quantum systems $12|3|\cdots|N$, and $\rho_{\alpha}$ are the corresponding reduced density matrices. According to the relation $ 1-Tr(\rho^2_{i_1~i_2\cdots i_k})= 1-Tr(\rho^2_{1 \cdots \widehat{i_1}\cdots \widehat{i_2} \cdots \widehat{i_k}\cdots N}),$
we have
\begin{eqnarray}\label{sec4-thm11}
 \nonumber
  \mathcal{C}_{N-1}^{2}(\rho_{12|3|\cdots|N}) &=& 2^{3-(N-1)}[(1-tr\rho_{12}^2)+\sum\limits_{i=3}^N(1-tr\rho_{i}^2)+
\sum\limits_{i=3}^N(1-tr\rho_{12i}^2)+\\
\nonumber
 && \sum\limits_{3 \leq i_1<i_2\leq N}(1-tr\rho_{i_1i_2}^2)+\cdots+\sum\limits_{3\leq i_1<i_2<\cdots<i_{\frac{N}{2}-2}\leq N}(1-tr\rho_{12i_1i_2\cdots i_{\frac{N}{2}-2}}^2)\\
   && +\sum\limits_{3\leq i_1<i_2<\cdots<i_{\frac{N}{2}-1}\leq N}(1-tr\rho_{i_1i_2\cdots i_{\frac{N}{2}-1}}^2)].
\end{eqnarray}

Concerning the number of the terms in the summations of (\ref{sec4-thm11}), we only need to consider the following non-trivial index $\alpha$, see \textbf{Table \ref{sec4-table1}}.
\begin{table}[htb]
\centering
\begin{tabular}{|c|c|}
\hline
  Details of $\alpha$ & Number of $\alpha$  \\ \hline
$12; ~3,4, \cdots, N$ & $C^{1}_{N-1}=C^{0}_{N-2}+C^{1}_{N-2}$  \\ \hline
$123,124, \cdots, 12N;~ 34, 35, \cdots, (N-1)N$ & $C^{2}_{N-1}=C^{1}_{N-2}+C^{2}_{N-2}$ \\ \hline
 $\cdots$ & $\cdots$ \\ \hline
\thead{$123\cdots \frac{N}{2},  \cdots, 12(\frac{N}{2}+2)\cdots N$ \\[1.5ex]
  $34\cdots(\frac{N}{2}+1),  \cdots, (\frac{N}{2}+2)(\frac{N}{2}+3)\cdots N$ } & $C^{\frac{N}{2}-1}_{N-1}=C^{\frac{N}{2}-2}_{N-2}+C^{\frac{N}{2}-1}_{N-2}$ \\ \hline
\end{tabular}
\textbf{\caption{Non-trivial index $\alpha$ in the equality (\ref{sec4-thm11}).}\label{sec4-table1}}
\end{table}
From equality (\ref{1-Tr}), we have that each item of $1-tr(\rho^2_{i_1i_2\cdots i_k})$ has the form $2|a_i|^2|a_j|^2$. To compute $\mathcal{C}_{N-1}^{2}(\rho_{12|3|\cdots|N})$ we just need to determine the coefficients $c_{ij}$ of the term $|a_i|^2|a_j|^2$ for each $1-tr(\rho^2_{i_1i_2\cdots i_k})$ in equality (\ref{sec4-thm11}).

%From \textbf{Lemma \ref{lem2}}, we can see that $C^{1}_{N-1}+\cdots+C^{\frac{N}{2}-1}_{N-1}=2^{N-2}-1$.
%\begin{equation}\label{sec4-thm1}
% \mathcal{C}_{N-1}^{2}(\rho_{12|3|\cdots|N}) = 2^{3-(N-1)}\sum\limits_{ 1\leq i<j\leq N}c_{ij}|a_i|^2|a_j|^2.
%\end{equation}
As $c_{12}=0$, we next calculate the coefficients $c_{ij}$ for $1\leq i<j\leq N$ and $(i,j)\neq (1,2)$. Taking the coefficient $c_{13}$ as an example, we can see from \textbf{Table \ref{N-1-N even}} that the total coefficient $c_{13}$ of the term $|a_1|^2|a_3|^2$ is $ 2^{3-(N-1)}(2\cdot 2C^{0}_{N-3}+2\cdot 2C^{1}_{N-3}+\cdots+2\cdot C^{\frac{N-2}{2}-1}_{N-3})$, which is equal to $4$ by \textbf{Lemma \ref{lem2}}.
\begin{table}[htb]
\centering
\begin{tabular}{|c|c|}
\hline
$1-Tr(\rho^2_{\alpha})$ containing the item $|a_1|^2|a_3|^2$  & The coefficient $c_{13}$ \\
   \hline
$1-Tr(\rho^2_{12})$; ~~$1-Tr(\rho^2_{3})$   & $2\cdot 2C^{0}_{N-3}$ \\
\hline
\thead{ $1-Tr(\rho^2_{124})$, $\cdots$,  $1-Tr(\rho^2_{12N})$ \\[2ex]
 $1-Tr(\rho^2_{34}), \cdots $ $1-Tr(\rho^2_{3N})$} & $2\cdot 2C^{1}_{N-3}$ \\
\hline
\thead{$1-Tr(\rho^2_{1245})$, $\cdots$,  $1-Tr(\rho^2_{12(N-1)N})$  \\[2ex]
 $1-Tr(\rho^2_{345}), \cdots $ $1-Tr(\rho^2_{3(N-1)N})$} & $2\cdot 2C^{2}_{N-3}$ \\
  \hline
  $\cdots$ & $\cdots$ \\
\hline
\thead{$1-Tr(\rho^2_{1245\cdots (\frac{N}{2}+1)})$, $\cdots$, $1-Tr(\rho^2_{12(\frac{N}{2}+2)\cdots N})$ \\[2ex]
  $1-Tr(\rho^2_{345\cdots (\frac{N}{2}+1)}), \cdots $ $1-Tr(\rho^2_{3(\frac{N}{2}+2)\cdots N})$} & $2\cdot 2C^{\frac{N-2}{2}-1}_{N-3}$ \\
\hline
\end{tabular}
\centering
\textbf{\caption{The coefficients of $|a_1|^2|a_3|^2$ for even $N$).}\label{N-1-N even}}
\end{table}
Other coefficients $c_{ij}$, $1\leq i<j\leq N$, $(i,j)\neq (1,2)$, can be calculated in a similar way, which are all equal to $4$. Hence we have $\mathcal{C}_{N-1}^{2}(\rho_{12|3|\cdots|N})=4\sum\limits_{\substack{ 1\leq i<j\leq N\\(i,j)\neq (1,2)}}|a_i|^2|a_j|^2$.

ii) $N\geq 3$ and $N$ is odd. %let $\rho_{12\cdots N} = |W\rangle_{A_1 A_2 \cdots A_N}\langle W|$, then by equality (\ref{M-partite}), we have
%\begin{equation}\label{sec4-2}
 %\mathcal{C}^2_{N-1}(\rho_{12|3|\cdots|N}) = 2^{2-(N-1)}[(2^{N-1}-2)-\sum\limits_{\alpha} Tr(\rho_{\alpha}^{2})],
%\end{equation}
%where the index $\alpha$ labels all $2^{N-1}-2$ non-trivial subsystems of the $(N-1)$-partite quantum systems $12|3|\cdots|N$ and $\rho_{\alpha}$ are the corresponding reduced density matrices.
%Associated with the conclusion $ 1-Tr(\rho^2_{i_1~i_2\cdots i_k})= 1-Tr(\rho^2_{1 \cdots \widehat{i_1}\cdots \widehat{i_2} \cdots \widehat{i_k}\cdots N}),$
Similarly we have
\begin{eqnarray}\label{sec4-thm31}
 \nonumber
  \mathcal{C}_{N-1}^{2}(\rho_{12|3|\cdots|N}) &=& 2^{3-(N-1)}[(1-tr\rho_{12}^2)+\sum\limits_{i=3}^N(1-tr\rho_{i}^2)+
\sum\limits_{i=3}^N(1-tr\rho_{12i}^2)+\\
\nonumber
 && \sum\limits_{3 \leq i_1<i_2\leq N}(1-tr\rho_{i_1i_2}^2)+\cdots+\sum\limits_{3\leq i_1<i_2<\cdots<i_{\frac{N-3}{2}}\leq N}(1-tr\rho_{12i_1i_2\cdots i_{\frac{N-3}{2}}}^2)\\
   && +\sum\limits_{3\leq i_1<i_2<\cdots<i_{\frac{N-1}{2}}\leq N}(1-tr\rho_{i_1i_2\cdots i_{\frac{N-1}{2}}}^2)].
\end{eqnarray}
The non-trivial index $\alpha$ we need to consider is shown in  \textbf{Table \ref{sec4-table2}}.
\begin{table}[htb]
\centering
\begin{tabular}{|c|c|}
\hline
  Details of $\alpha$ & Number of $\alpha$  \\ \hline
$12; ~3,4, \cdots, N$ & $C^{1}_{N-1}=C^{0}_{N-2}+C^{1}_{N-2}$  \\ \hline
$123,124, \cdots, 12N;~ 34, 35, \cdots, (N-1)N$ & $C^{2}_{N-1}=C^{1}_{N-2}+C^{2}_{N-2}$ \\ \hline
 $\cdots$ & $\cdots$ \\ \hline
 \thead{ $123\cdots \frac{N-1}{2},  \cdots, 12(\frac{N-1}{2}+4)\cdots N $ \\ [2ex]
  $ 34\cdots(\frac{N-1}{2}+1),  \cdots, (\frac{N-1}{2}+2)(\frac{N-1}{2}+3)\cdots N$ } & $C^{\frac{N-1}{2}-1}_{N-1}=C^{\frac{N-1}{2}-2}_{N-2}+C^{\frac{N-1}{2}-1}_{N-2}$ \\
\hline
$123\cdots (\frac{N-1}{2}+1),  \cdots, 12(\frac{N-1}{2}+3)\cdots N $ & $\frac{1}{2}C^{\frac{N-1}{2}}_{N-1}=C^{\frac{N-1}{2}-1}_{N-2}$ \\
\hline
\end{tabular}
\textbf{\caption{Non-trivial index $\alpha$ in the equality (\ref{sec4-thm31}).}\label{sec4-table2}}
\end{table}

%From \textbf{Lemma \ref{lem1}}, we can see that $C^{1}_{N-1}+\cdots+\frac{1}{2}C^{\frac{N-1}{2}}_{N-1}=2^{N-2}-1$. Combing equalities (\ref{1-Tr}) and (\ref{sec4-thm31}),  hence we have
%\begin{eqnarray}\label{sec4-thm3}
%  \mathcal{C}_{N-1}^{2}(\rho_{12|3|\cdots|N}) &=&  2^{3-(N-1)}\sum\limits_{ 1\leq i<j\leq N}c_{ij}|a_i|^2|a_j|^2.
%\end{eqnarray}
To compute $\mathcal{C}_{N-1}^{2}(\rho_{12|3|\cdots|N})$ we need to determine the
coefficients $c_{ij}$ of the term $|a_i|^2|a_j|^2$ for each $1-tr(\rho^2_{i_1i_2\cdots i_k})$ in equality (\ref{sec4-thm31}). Obviously, $c_{12}=0$. We still take the coefficient $c_{13}$ as an example. From \textbf{Table \ref{N-1-N odd}} we have that the total coefficient $c_{13}$ of the term $|a_1|^2|a_3|^2$ is $ 2^{3-(N-1)}(2\cdot 2C^{0}_{N-3}+2\cdot 2C^{1}_{N-3}+\cdots+2\cdot 2C^{\frac{N-1}{2}-2}_{N-3}+2\cdot C^{\frac{N-1}{2}-1}_{N-3})$, which is equal to $4$ by \textbf{Lemma \ref{lem1}}.
\begin{table}[htb]
\centering
\begin{tabular}{|c|c|}
\hline
$1-Tr(\rho^2_{\alpha})$ which contains the item $|a_1|^2|a_3|^2$  & The coefficients $c_{13}$ \\
   \hline
$1-Tr(\rho^2_{12})$; ~~$1-Tr(\rho^2_{3})$   & $2\cdot 2C^{0}_{N-3}$ \\
\hline
\thead{ $1-Tr(\rho^2_{124})$, $\cdots$,  $1-Tr(\rho^2_{12N})$ \\[2ex]
 $1-Tr(\rho^2_{34}), \cdots $ $1-Tr(\rho^2_{3N})$} & $2\cdot 2C^{1}_{N-3}$ \\
\hline
\thead{$1-Tr(\rho^2_{1245})$, $\cdots$,  $1-Tr(\rho^2_{12(N-1)N})$  \\[2ex]
 $1-Tr(\rho^2_{345}), \cdots $ $1-Tr(\rho^2_{3(N-1)N})$} & $2\cdot 2C^{2}_{N-3}$ \\
  \hline
  $\cdots$ & $\cdots$ \\
\hline
\thead{$1-Tr(\rho^2_{124\cdots (\frac{N-1}{2}+1)})$, $\cdots$, $1-Tr(\rho^2_{12(\frac{N-1}{2}+4)\cdots N})$ \\[2ex]
  $1-Tr(\rho^2_{34\cdots (\frac{N-1}{2}+1)}), \cdots $ $1-Tr(\rho^2_{3(\frac{N-1}{2}+4)\cdots N})$} & $2\cdot 2C^{\frac{N-1}{2}-2}_{N-3}$ \\
\hline
$1-Tr(\rho^2_{124\cdots (\frac{N-1}{2}+2)})$, $\cdots$, $1-Tr(\rho^2_{12(\frac{N-1}{2}+3)\cdots N})$ & $ 2\cdot 1C^{\frac{N-1}{2}-1}_{N-3}$ \\
\hline
\end{tabular}
\centering
\textbf{\caption{The coefficients of $|a_1|^2|a_3|^2$ ($N$ odd).}\label{N-1-N odd}}
\end{table}
Similarly, we can prove that other coefficients $c_{ij}$, $1\leq i<j\leq N$, $(i,j)\neq (1,2)$ are all equal to $4$.
Therefore, $\mathcal{C}_{N-1}^{2}(\rho_{12|3|\cdots|N})=4\sum\limits_{\substack{ 1\leq i<j\leq N\\(i,j)\neq (1,2)}}|a_i|^2|a_j|^2$.  $\Box$

\begin{corollary}\label{N-1  cor}
For the $N$-qubit $W$-class states $\rho_{12 \cdots N} = |W\rangle_{A_1 A_2 \cdots A_N}\langle W|$, we have
\begin{equation}\label{}
\mathcal{C}_{N-1}^2(\rho_{ij|1|\cdots|\widehat{i}|\cdots|\widehat{j}|\cdots|N})
=4\sum\limits_{\substack{ 1\leq k<l\leq N\\(k,l)\neq (i,j)}}|a_k|^2|a_l|^2
\end{equation}
for $N\geq 3$ and $1\leq i<j\leq N$,
\end{corollary}

Based on the above conclusions, the relation between the $N$-partite concurrence and the $(N-1)$-partite concurrence for $W$-class states are given by the following theorem.

\begin{theorem}\label{theorem5}
For the $N$-qubit $W$-class states $|W\rangle_{A_1 A_2 \cdots A_N}$, we have
\begin{equation}\label{}
\mathcal{C}_{N}^2(\ket{W_{A_1 A_2 \cdots A_N}})= \frac{1}{C_N^2-1}\widetilde{\mathcal{C}}_{N-1}^2(\ket{W_{A_1 A_2 \cdots A_N}})
\end{equation}
for $N\geq 3$.
\end{theorem}

\noindent{\bf Proof.}
From \textbf{Corollary \ref{N-1 cor}}, we have
\begin{eqnarray}\label{}
 \nonumber
   \widetilde{\mathcal{C}}_{N-1}^2(\ket{W_{A_1 A_2 \cdots A_N}})&=&\sum\limits_{\mathcal{P}}\mathcal{C}_{N-1}^2(\rho_{\mathcal{P}}) \\
\nonumber
   &=&4\cdot (C_N^2-1)\sum\limits_{ 1\leq i<j\leq N}|a_i|^2|a_j|^2,
\end{eqnarray}
where the index ${\mathcal{P}}$ labels all $C_N^2$ different $(N-1)$-partite partitions of $N$-partite systems.
By \textbf{Theorem \ref{Sec2-3}}, we have $\mathcal{C}_{N}^2(\ket{W_{A_1 A_2 \cdots A_N}})= \frac{1}{C_N^2-1}\widetilde{\mathcal{C}}_{N-1}^2(\ket{W_{A_1 A_2 \cdots A_N}})$. $\Box$

\section{Generalized results to get lower bound of multipartite concurrence}\label{sec5}
\textbf{Theorem \ref{theorem5}} gives an explicit expression of the concurrence for pure multipartite $W$-class states. Based on \textbf{Theorem \ref{theorem5}} we can also derive tighter lower bounds of concurrence for mixed multipartite states.

Let us consider the case of $N=4$. We first take a look at the $4$-qubit $W$-class states $ |W\rangle_{A_1 A_2 A_3 A_4}$ with density matrix $\rho_{1234}=|W\rangle_{A_1 A_2 A_3 A_4}\langle W|$. From equality (\ref{14}) we have
\begin{equation}\label{sec4-1}
  \mathcal{C}_4^{2}({|W\rangle}_{A_1 A_2 A_3 A_4})=\frac{1}{2}[\sum\limits_{i=1}^4(1-tr\rho_{i}^2)+\sum\limits_{i=2}^4(1-tr\rho_{1i}^2)].
\end{equation}
%$\mathcal{C}_N^{2}({|W\rangle}_{A_1 A_2 A_3 A_4})=\frac{1}{2}[\sum\limits_{i=1}^4(1-tr\rho_{i}^2)+\sum\limits_{i=2}^4(1-tr\rho_{1i}^2)]$.
From equality (\ref{sec4-thm11}) we get
$\mathcal{C}_{3}^{2}(\rho_{12|3|4})=(1-Tr\rho_{12}^2)+(1-Tr\rho_{3}^2)+(1-Tr\rho_{4}^2)$. Therefore,
\begin{equation}\label{sec4-2}
\widetilde{\mathcal{C}}_{3}^2(\ket{W_{A_1 A_2 A_3 A_4}})=3\sum\limits_{i=1}^4(1-tr\rho_{i}^2)+2\sum\limits_{i=2}^4(1-tr\rho_{1i}^2).
\end{equation}
%$\widetilde{\mathcal{C}}_{3}^2(\ket{W_{A_1 A_2 A_3 A_4}})=3\sum\limits_{i=1}^4(1-tr\rho_{i}^2)+2\sum\limits_{i=2}^4(1-tr\rho_{1i}^2)$.
Moreover, from \textbf{Theorem \ref{theorem5}}, we obtain $\mathcal{C}_4^{2}({|W\rangle}_{A_1 A_2 A_3 A_4})=\frac{1}{5}\widetilde{\mathcal{C}}_{3}^2(\ket{W_{A_1 A_2 A_3 A_4}})$.
Hence, for the $4$-qubit $W$-class states $|W\rangle_{A_1 A_2 A_3 A_4}$ we have
$\sum\limits_{i=1}^4(1-tr\rho_{i}^2)=\sum\limits_{i=2}^4(1-tr\rho_{1i}^2)$. Under such particular properties, we have the following lower bounds for some multipartite mixed states.

\begin{theorem}\label{theorem7}
For any $4$-partite quantum state $\rho \in  H_1\otimes H_2\otimes H_3 \otimes H_4,$ if $\rho = \sum\limits_{i}p_{i}|\psi_{i}\rangle\langle\psi_{i}|$ attains the minimal partition of the multipartite concurrence and $\sum\limits_{i=1}^4(1-tr\rho_{i}^2)=\sum\limits_{i=2}^4(1-tr\rho_{1i}^2)$ for any $|\psi_{i}\rangle$ in above partition, then
\begin{equation}\label{lower bound2}
C_4^{2}(\rho) \geq  \frac{1}{5}\widetilde{C_3}^{2} (\rho),
\end{equation}
where $\widetilde{C_3}^{2} (\rho) = C_3^{2}(\rho_{1|2|34})+C_3^{2}(\rho_{1|3|24})+C_3^{2}(\rho_{1|4|23})+C_3^{2}(\rho_{12|3|4})+C_3^{2}(\rho_{13|2|4})+C_3^{2}(\rho_{14|2|3}).$
\end{theorem}

\noindent{\bf Proof.}
First consider the pure state $|\psi\rangle \in H_1\otimes H_2\otimes H_3 \otimes H_4$ with $\rho=|\psi\rangle\langle\psi|$. From equality (1) we have
\begin{eqnarray}\label{26}
C_4^{2}(\rho)=\frac{1}{2} (\sum_{i=1}^{4}(1-tr\rho^{2}_{i}) + \sum_{i=2}^{4}(1-tr\rho^{2}_{1i}))
\end{eqnarray}
and
\begin{eqnarray}\label{27}
C_3^{2}(\rho_{i|j|kl})= (1-tr\rho^{2}_{i}) + (1-tr\rho^{2}_{j}) + (1-tr\rho^{2}_{kl}),
\end{eqnarray}
where $\rho_{i}=Tr_{jkl}(\rho)$, $\rho_{j}=Tr_{ikl}(\rho)$ and $\rho_{kl}=Tr_{ij}(\rho).$
Since $\sum\limits_{i=1}^4(1-tr\rho_{i}^2)=\sum\limits_{i=2}^4(1-tr\rho_{1i}^2)$ we have $C_4^{2}(\rho)=\sum\limits_{i=1}^{4}(1-tr\rho^{2}_{i})$ and  $\widetilde{C_3}^{2} (\rho) = 5(\sum\limits_{i=1}^{4}(1-tr\rho^{2}_{i})),$ hence we get
$C_4^{2}(\rho) =  \frac{1}{5}\widetilde{C_3}^{2} (\rho)$, where $\widetilde{C_3}^{2} (\rho) = C_3^{2}(\rho_{1|2|34})+C_3^{2}(\rho_{1|3|24})+C_3^{2}(\rho_{1|4|23})
+C_3^{2}(\rho_{12|3|4})+C_3^{2}(\rho_{13|2|4})+C_3^{2}(\rho_{14|2|3}).$

As a mixed state $\rho = \sum_{i}p_{i}|\psi_{i}\rangle\langle\psi_{i}|$ attains the minimal partition of the multipartite concurrence, and $\sum\limits_{i=1}^4(1-tr\rho_{i}^2)=\sum\limits_{i=2}^4(1-tr\rho_{1i}^2)$ for any $|\psi_{i}\rangle$ in above partition, we have
$$C_{4}^{2}(\rho) = (\sum_{i}p_{i}C_{4}(|\psi_{i}\rangle\langle\psi_{i}|))^{2}~~~~~~~~~~~~~~~~~~~~~~~~~~~~~~~~~~~~~~~~~~~~~~~~~~~~~~~~~~~~~~~~~~~~~~~~~$$
$$~~~ = (\sum_{i}p_{i}\sqrt{\frac{1}{5}(C_3^{2}((|\psi_{i}\rangle)_{1|2|34})
+C_3^{2}((|\psi_{i}\rangle)_{1|3|24})+\cdots
+C_3^{2}((|\psi_{i}\rangle)_{14|2|3}))}~)^{2}~~~~~~~~~~~~~~~~~~~~~~~~$$
$$~~\geq (\sum_{i}p_{i}\frac{1}{\sqrt{6}}C_3((|\psi_{i}\rangle)_{1|2|34}))^{2} + (\sum_{i}p_{i}\frac{1}{\sqrt{5}}C_3((|\psi_{i}\rangle)_{1|3|24}))^{2} + \cdots + (\sum_{i}p_{i}\frac{1}{\sqrt{5}}C_3((|\psi_{i}\rangle)_{14|2|3}))^{2}$$
$$\geq \frac{1}{5}(C_3^{2}(\rho_{1|2|34})+C_3^{2}(\rho_{1|3|24})+C_3^{2}(\rho_{1|4|23})+C_3^{2}(\rho_{12|3|4})+C_3^{2}(\rho_{13|2|4})+C_3^{2}(\rho_{14|2|3})),~~~~~~~~$$
\noindent where the relation $(\sum_{j}(\sum_{i}x_{ij})^{2})^{\frac{1}{2}}\leq \sum_{i}(\sum_{j}x_{ij}^{2})^{\frac{1}{2}}$ has been used in first inequality. Therefore, we get (\ref{lower bound2}). ~~~~~~~~~~~~~~~~~~~~~~~~~~~~~~~~~~~~~~~~~~~~~~~~~~~~~~~~~~~~~~~~~~~~~~~~~~~~~~~~~~~~~~~~~~$\Box$

To illustrate our lower bound (\ref{lower bound2}), let us consider the following example.

\begin{Exa}\label{ex4qubit}
Consider the $4$-qubit $W$-class state
$\ket W_{A_1 A_2 A_3 A_4} = a_1 \ket {1000} +a_2 \ket{0100} + a_3 |0010\rangle +a_4 |0001\rangle ,$
where $\sum\limits_{i=1}^{4}|a_i|^2=1$.
Denote $\rho=|W\rangle_{A_1 A_2 A_3 A_4}\langle W|$. We have $\mathcal{C}_4^{2}(\rho) =\frac{1}{2}[(1-tr(\rho_1^2))+(1-tr(\rho_2^2))+(1-tr(\rho_3^2))
+(1-tr(\rho_4^2))+(1-tr(\rho_{12}^2))+(1-tr(\rho_{13}^2))+(1-tr(\rho_{14}^2))]$ and
$1-tr(\rho_1^2)=2|a_1|^2(|a_2|^2+|a_3|^2+|a_4|^2)$. Similarly we have
$$1-tr(\rho_2^2)=2|a_2|^2(|a_1|^2+|a_3|^2+|a_4|^2), 1-tr(\rho_3^2)=2|a_3|^2(|a_1|^2+|a_2|^2+|a_4|^2),$$
$$1-tr(\rho_4^2)=2|a_4|^2(|a_1|^2+|a_2|^2+|a_3|^2), 1-tr(\rho_{12}^2)=2(|a_1|^2+|a_2|^2)(|a_3|^2+|a_4|^2),$$
$$1-tr(\rho_{13}^2)=2(|a_1|^2+|a_3|^2)(|a_2|^2+|a_4|^2), 1-tr(\rho_{14}^2)=2(|a_1|^2+|a_4|^2)(|a_2|^2+|a_3|^2).$$
We easily verify that $\sum\limits_{i=1}^4(1-tr\rho_{i}^2)=\sum\limits_{i=2}^4(1-tr\rho_{1i}^2)$.
Hence by \textbf{Theorem \ref{theorem7}} we can get $\mathcal{C}_4^2(\ket{W_{A_1 A_2 A_3 A_4}})=\frac{1}{5}\widetilde{\mathcal{C}}_3^2(\ket{W_{A_1 A_2 A_3 A_4}}),$ which is better than the lower bound given in \cite{Ch1}.
\end{Exa}

\section{Conclusions}\label{sec6}
The multipartite concurrence plays important roles in quantifying the entanglement of multipartite quantum systems. By taking into account the structures of $W$-class states we have studied the multipartite concurrence for arbitrary multipartite $W$-class states in terms of the $(N-1)$-partitions of subsystems. We have used the method of permutation and combination to present explicit expressions of multipartite concurrence for arbitrary $W$-class states.
We have shown the relations between the multipartite concurrence and the $(N-1)$-partite concurrence for $W$-class states. At last, motivated by the explicit expressions of concurrence for $W$-class states, we have presented a lower bound of concurrence for a class of four-partite mixed states. Similarly, our approach may be also applied to study the multipartite concurrence for arbitrary $N$-partite $W$-class states based on arbitrary $M$-partite partition of subsystems, and to give relations between the multipartite concurrence and arbitrary $M$-partite ($2\leq M \leq N-2$) concurrence for $W$-class states, as well as the corresponding lower bounds of concurrence for more general multipartite mixed states.

\vspace{2.5ex}
\noindent{\bf Acknowledgments}\, \,
This work was supported by the Basic and Applied Basic Research Funding Program of Guangdong Province (Grant No. 2019A1515111097), Yunnan Provincial Research Foundation for Basic Research, China (Grant No. 202001AU070041) and Guangdong Universities' Special Projects in Key Fields of Natural Science (No.2019KZDZX1005), National Natural Science Foundation of China (NSFC) under Grants 12075159 and 12171044, Beijing Natural Science Foundation (Grant No. Z190005), and the Academician Innovation Platform of Hainan Province.

\end{document}